# VALIDITY IN DESIGN SCIENCE

**Kai R. Larsen**
Leeds School of Business, University of Colorado, Boulder, CO 80309 U.S.A.
{Kai.Larsen@colorado.edu}

**Roman Lukyanenko**
McIntire School of Commerce, University of Virginia, Charlottesville, VA 22908 U.S.A.
{romanl@virginia.edu}

**Roland M. Mueller**
Department of Business and Economics, Berlin School of Economics and Law, Berlin, Germany
{roland.mueller@hwr-berlin.de}

**Veda C. Storey**
J. Mack Robinson College of Business, Georgia State University, Atlanta, GA 30303 U.S.A.
{vstorey@gsu.edu}

**Jeffrey Parsons**
Faculty of Business Administration, Memorial University of Newfoundland, St. John's, NL Canada
{jeffreyp@mun.ca}

**Debra VanderMeer**
College of Business, Florida International University, Miami, FL 33199 U.S.A. {vanderd@fiu.edu}

**Dirk S. Hovorka**
University of Sydney Business School, University of Sydney, Sydney, Australia
{dirk.hovorka@sydney.edu.au}

*Researchers must ensure that the claims about the knowledge produced by their work are valid. However, validity is neither well-understood nor consistently established in design science, which involves the development and evaluation of artifacts (models, methods, instantiations, and theories) to solve problems. As a result, it is challenging to demonstrate and communicate the validity of knowledge claims about artifacts. This paper defines validity in design science and derives the Design Science Validity Framework and a process model for applying it. The framework comprises three high-level claim and validity types—criterion, causal, and context—as well as validity subtypes. The framework guides researchers in integrating validity considerations into projects employing design science and contributes to the growing body of research on design science methodology. It also provides a systematic way to articulate and validate the knowledge claims of design science projects. We apply the framework to examples from existing research and then use it to demonstrate the validity of knowledge claims about the framework itself.*

**Keywords**: design science, design science research (DSR), Design Science Validity Framework, knowledge claim, research validity, criterion validity, causal validity, context validity, characteristic validity, efficacy validity, external validity, ecological validity, evaluation, validation

## Introduction

Scientific knowledge must be credible and dependable (Burton-Jones et al. 2021; Creswell and Miller 2000). An established validity tradition facilitates credible and dependable science because it provides a systematic way of evaluating the claimed contributions to knowledge in research studies (Cook and Campbell 1979; Guba and Lincoln 1994). Each scientific discipline has patterns and procedures to evaluate the validity of knowledge claims regarding phenomena of interest. Articulating these patterns provides evaluative clarity, increases research efficiency, promotes the sharing of best practices, facilitates cumulative science, and contributes to greater public trust in science.

In *design science* – a genre of research in information systems, computer science, engineering, medicine, management, and material and biomedical sciences – researchers advance knowledge claims about the characteristics or performance of the artifacts they invent. Researchers have proposed processes for conducting design science, each of which identifies evaluation as an essential step (Gregor and Hevner 2013; Hevner et al. 2004; Peffers et al. 2007; Prat et al. 2015; Sein et al. 2011). However, existing methods and frameworks neither consider the validity of knowledge claims to be a core component of artifact evaluation nor agree on what it means to assess the validity of knowledge claims about artifacts.

Design science exhibits recurring patterns of knowledge claims. Examples include: the artifact outperforms the state of the art (Padmanabhan et al. 2022); a recent version of the artifact performs better than the previous version(s) (Sein et al. 2011); the model corresponds to a referent real-world system (Gonzalez-Huerta et al. 2017); and the artifact works due to the presence of particular design features (Abbasi et al. 2012). There is also a repertoire of common evaluation approaches in design science (Prat et al. 2015; Venable et al. 2016), although these do not address the logic that connects claims to validation, the process of establishing validity of a particular type of knowledge claim.

While validity types are used in some design science projects, these are generally adapted from other traditions. For example, internal validity, and external validity are adapted from psychometrics (Baskerville et al. 2015; vom Brocke, Hevner, et al. 2020), whereas validity metrics, such as precision and recall, are taken from information retrieval, statistics, and other disciplines (Larsen and Becker 2020). An exception is instantiation validity (Lukyanenko and Parsons 2020), which was proposed as a native design science validity type.

Most design science validations lack any label, making it difficult to reference them effectively and share best practices with the community. Furthermore, some evaluation procedures, such as applicability checks (Rosemann and Vessey 2008), respond to multiple (often unstated) claims, leading to questions regarding *what exactly* has been validated. Despite the potential benefits, there have been no systematic attempts to define or survey validity across all the artifact types addressed by design scientists (e.g., models, methods, design theories; see Gregor and Hevner 2013) or provide a process for validating claims. Hence, the essential question of what constitutes validity in design science remains unanswered.

A systematic approach to validity will benefit design science. First, it will promote high-quality scholarship. Critical to the quality of design science is the link between knowledge claims and the evidence that supports these claims. This can be facilitated by following evaluation procedures in accordance with agreed-upon structures known to support the types of claims made. These procedures commonly demand procedural rigor and careful argumentation (e.g., evidence of causality for internal validity, often provided through randomized experiments). Establishing validity would make the linkage between claims and their validation more explicit and more amenable to attention and scrutiny. This transparency is important for engendering trust and confidence in the claims and reported findings (Burton-Jones et al. 2021).

Second, as specific types of validity address specific types of claims (e.g., internal validity addresses claims of causation), establishing a validity tradition for design science will lead to making knowledge claims more explicit. This can help researchers better appreciate and communicate the contributions they are making and facilitate the validation process. Furthermore, the explicated knowledge claims can also be used by other researchers seeking to extend the original contributions both within and beyond design science, thus allowing design scientists to better contribute to a cumulative tradition of research on the design, use, and impact of information technology artifacts.

Third, a comprehensive conceptualization of validity will increase the impact of design science on the real world. By sensitizing researchers to the nature of knowledge claims and validity types, artifacts that fail to serve their intended purpose when exposed to unexpected real-world circumstances are less likely. For example, in machine learning it is important for researchers to be aware of which artifact features produced the improved outcomes. Still, these are narrowly focused on the performance of a proposed artifact. A more inclusive conception of validity would consider an artifact's applicability for varied real world conditions, including for broader tasks and different user groups (Ethayarajh and Jurafsky 2020).

Fourth, establishing a validity tradition will facilitate the publication of design science studies. As a reflection of community norms and shared concepts, a validity tradition boosts research productivity by increasing validation consistency (Chan 2014) and streamlining validation practices. Note that the prolific fields of econometrics and psychometrics have distinct validity or validation approaches (Taylor 2013), as do qualitative research (Creswell and Miller 2000) and areas of computer science and software engineering, such as model simulation (e.g., Pääkkönen et al. 2017) and problem decidability (e.g., Fan et al. 2018). Across disciplines, a higher level of agreement on core disciplinary norms (including validation) correlates with increased research productivity and publication quality (Gumport 2007). The formalization of shared disciplinary conceptualizations is vital to accelerating scientific progress (National Academies of Sciences 2022).

Finally, establishing the nature of validity in design science will position it with respect to established traditions (e.g., in the behavioral sciences) and better communicate the distinct contributions of design science to outsiders. The common patterns of validation receive a systematic name, thereby helping design science scholars and outsiders reference these evaluation approaches. This, in turn, contributes to establishing design science identity.

This research makes several contributions. We first examine how validity concepts have been used in design science. We then consider the general nature of validity to identify the established foundations that could benefit design scientists. We do so by developing insights into the logic and relationships among the claims, artifact types, validity types, and evaluation context. These foundations then allow us to define validity types and use them appropriately to evaluate design science knowledge claims.

Next, we develop the Design Science Validity Framework to provide a systematic way to articulate and validate the knowledge claims of design science projects. The framework comprises three high-level knowledge claim and validity types—*criterion, causal,* and *context*—as well as validity subtypes. This structure corresponds to the aim of design science to develop innovative artifacts as solutions to societal challenges, while contributing scientific knowledge that practitioners can reuse in diverse contexts. The framework is inclusive in its coverage of design science artifacts, including implemented systems (e.g., tools and deployed instantiations), abstract contributions that encapsulate blueprints for systems development (e.g., conceptual models or machine learning methods), and theoretical design knowledge (e.g., design theories that prescribe design and action to attain specific goals, see Gregor and Jones 2007).

Finally, we evaluate the Design Science Validity Framework through a series of studies informed recursively by the framework itself. Through two applicability checks, an extensive analysis of research on validity, and a comprehensive analysis of design science evaluation, we demonstrate the utility of our framework for researchers, its ability to capture existing validation practices, and its parsimony. We conclude with a general discussion on the nature and importance of design science validity, recommend how the framework can be used, and identify future research opportunities.

## METHODOLOGICAL DEVELOPMENTS IN DESIGN SCIENCE

Design science has developed methodological foundations to facilitate credible and dependable practical scientific knowledge. These efforts have focused on ensuring rigor in artifact development and evaluating the utility and quality of the artifacts and design knowledge. (See Table 1 for specific topics pursued). Whereas rigor in artifact development has received much attention, rigor in evaluating the utility and quality of artifacts and design knowledge has lagged, even though this is a major challenge.

**Table 1. Methodological Foundations in Design Science**

| Contribution | Reference |
|---|---|
| Integration of systems development with rigorous theory and empirical evaluation | Nunamaker et al. (1991); March and Smith (1995) |
| Approaches to evaluating artifacts and theories | Hevner et al. (2004); Venable et al. (2016); Prat et al. (2015); Gregor and Jones (2007); Tuunanen et al. (2024) |
| Methods to produce and communicate design science | Peffers et al. (2007); Gregor and Hevner (2013); Johannesson and Perjons (2014); Baskerville et al. (2015); Avdiji and Winter (2019); Iivari (2015); Gregor et al. (2020); Tuunanen et al. (2024) |
| Alignment of kernel theories with artifacts | Arazy et al. (2010); Gregor and Jones (2007); Kuechler and Vaishnavi (2012); vom Brocke, Winter et al. (2020) |
| Transparency | vom Brocke et al. (2021); Burton-Jones et al. (2021); Lukyanenko and Parsons (2020); Hevner et al. (2024) |
| Alignment with practice | Sein et al. (2011); Lukyanenko and Parsons (2020); Hevner et al. (2024) |
| Design theorization | Mandviwalla (2015); Gregor and Jones (2007); Gregory and Muntermann (2014); Lukyanenko and Parsons (2020); Gregor et al. (2020) |

A significant step toward establishing rigor for design science is understanding what evaluations already exist, how they provide complementary knowledge, and how they overlap. No previous work has attempted to understand the rigor of design science evaluation by establishing the possible types of validities. Doing so requires understanding types of knowledge claims, artifacts, and validation approaches.

We reviewed prominent design science framework journal articles as summarized in Table 2. Most frameworks in Table 2 deal primarily with the rigor of artifact development. For example, Sein et al. (2011) focused on action design research (ADR), where evaluations occur in a real-world, authentic context. Venable et al. (2016) distinguished between formative and summative evaluations, focusing on when evaluations are conducted. Gregor and Hevner (2013) considered ways to position and present design science and whether evaluations occur inside or outside a development context. Prat et al. (2015) organized evaluation approaches into a taxonomy of five dimensions of distinct evaluation types. Baskerville et al. (2015) developed a framework to highlight the different types of knowledge production that can occur throughout a design science project. Tuunanen et al. (2024) provided an approach for managing complex, multi-stage projects by suggesting that these stages ("echelons") produce different artifacts that can be evaluated at each stage. It is noteworthy that validity concerns have grown over time from little to no consideration in early frameworks (e.g., Hevner et al. 2004; Peffers et al. 2007) to active consideration in recent methodological studies (e.g., Baskerville et al. 2015; Tuunanen et al. 2024).

The popular frameworks in Table 2 constitute the established methodological guidance in design science. They provide clarity for understanding the general approach of conducting research and evaluating artifacts but do not offer guidance on validation activities, including how evidence can be provided to justify knowledge claims.

**Table 2. Design Science Frameworks and the Role of Validity**

| Paper | Framework overview | Validity and claims | Evaluation |
|---|---|---|---|
| Hevner et al. (2004) | Design science consists of building and evaluating artifacts and connecting them to the environment and design knowledge base. | Minimal emphasis on validity. Cites other work on how the accumulation of evidence will eventually establish the validity of broader design science claims. | Proposes two fundamental questions that require evidence from evaluations: 1) what utility does the artifact provide? and 2) how is that utility demonstrated? Proposes that contribution arises from utility demonstrated through evaluation. |
| Peffers et al. (2007) | A six-step iterative process for presenting and evaluating design science. | Little emphasis on validity. Suggests that design science has had to employ "ad hoc arguments" (p. 50). | Researchers observe and measure how well the artifact supports a solution. Involves comparing objectives of a solution to actual observed results from using artifact (e.g., via response time, items produced, user satisfaction surveys, and simulation). |
| Sein, et al. (2011) | A four-stage model of action design research. | Contains no mention of validity or knowledge claims. | Argues for the principle of "authentic and concurrent evaluation" to emphasize a key characteristic of action design research. |
| Gregor and Hevner (2013) | A schema for publication in design science, including evaluation. | Artifact evaluation in terms of criteria that can include validity, utility, quality, and efficacy. | Distinguishes evaluation inside and outside the development environment. Utility criteria related to whether performance transfers outside of the development environment. |
| Prat et al. (2015) | A taxonomy of evaluation methods with five dimensions: goal, environment, structure, activity, and evaluation. | Validity is part of the goal dimension of efficacy and effectiveness. Validity is attained if artifact works correctly (achieves goal). | Focuses on "what" (systems of artifacts) and "how" (methods used) in evaluation. Evaluation involves the "relativeness" of the artifact's superiority to other solutions. Identifies typical evaluation techniques and secondary evaluations. |
| Baskerville et al. (2015) | Genres of inquiry. Framework: design vs. science and nomothetic vs. idiographic. | Provides 18 quality criteria. Mentions internal and external validity. | Provides reflections on genres of inquiry, rather than evaluation itself. |
| Venable et al. (2016) | Design science evaluations exist along two dimensions: artificial vs. naturalistic and formative vs. summative. | Validity is treated as emerging from the strength of evaluation and how well the artifact accomplishes the purpose. | Key purpose of evaluation is to assess how well an artifact achieves the intended utility. |

Some work in design science has recognized the importance of establishing validity (vom Brocke, Winter, et al. 2020) and suggested that existing behavioral validity types (e.g., internal, ecological) might apply to design science (Baskerville et al. 2015). In design science employing artificial intelligence-based techniques, such as machine learning and natural language processing, it is customary to report precision, recall, and the $F_1$-score, as well as other confusion matrix metrics (e.g., Abbasi and Chen 2008; Li et al. 2020). These are not types of validity but are validity-related because they provide quantifiable measures to establish the validity of claims. They also do not apply to all design science artifacts (e.g., design theories). To establish validity in design science, we first consider validity in science broadly, which we then synthesize with specific concerns of design science.[1]

## RESEARCH VALIDITY

The basic idea that knowledge claims should be validated dates to antiquity (Carter 2019). The term *validity* originated in the quantitative social sciences (e.g., Cronbach and Meehl 1955) and later became more widely used as it was applied to evolving cross-disciplinary beliefs about creating appropriate evaluations (e.g., in which prior knowledge is assessed, and the acceptance criteria are agreed upon) to establish the validity of knowledge claims.

To understand validity in design science, we reviewed approximately 7,500 sources from the literature on research validity from information systems (including design science) and other research areas, such as computer science, social sciences, engineering, medicine, law, and humanities. This effort yielded 2,418 candidate validities, constituting the largest effort to review validities by more than an order of magnitude. We report on these efforts later when we validate our framework. Here, we briefly review how validity is used in science, as this informs our notions of design science validity.

Initially, validity was narrowly understood as "the closeness of what we believe we are measuring to what we intended to measure" (Roberts and Priest 2006, p. 41). This view focused validity on measurement artifacts, especially in psychometrics, concerned with tests, instruments, or questionnaires administered to humans. Over time, quantitative researchers developed many different kinds of validity (e.g., internal, ecological, discriminant, external) now widely used in information systems behavioral research (Boudreau et al. 2001). Using these validities has been recognized as a "professional responsibility" (Shultz et al. 1998, p. 266).

Qualitative and interpretivist researchers have argued for unique validity concerns in their work, emphasizing the importance of the context, setting, and participants, in addition to the role of the researchers in creating a natural, trustworthy, confirmable, and dependable account of the research process (Lincoln and Guba 1985; Onwuegbuzie and Leech 2007). Validity in this tradition is considered to be a product of social consensus, where what is valid is based on a "community of acceptability" (Moules et al. 2015, p. 172). Reflecting on these efforts, Creswell and Miller (2000, p. 124) noted "a general consensus" that validity is foundational to the acceptance of studies – a view widely shared across disciplines (Hoyningen-Huene 2013), including information systems (Burton-Jones et al. 2021).

In computer science, a discipline closely related to design science in information systems, researchers in subfields such as machine learning and artificial intelligence have embraced a common task framework, focusing on shared benchmarks to validate new artifacts relative to the state-of-the-art as reflected in shared leaderboards (Matadamas-Hernández et al. 2012). This framework has catalyzed major advances in computer vision and natural language processing. For example, machine learning researchers have carefully developed approaches, such as causation analysis and ablation studies, to evaluate designed artifacts (e.g., Chowdary and Kanhangad 2022). They employ these approaches under the common task framework to test knowledge claims and contribute to the shared task knowledge. Nevertheless, despite impressive progress the common task framework may have limited the practical applications of resulting models and disadvantaged evaluation criteria such as "compactness, fairness, and energy efficiency" (Ethayarajh and Jurafsky 2020, p. 1).

[1] Appendix A evaluates the current state of validation practices.

Each type of inquiry, and sometimes each area (e.g., machine learning), has developed separate validity traditions. Just as qualitative and quantitative validity concerns differ, design science validity can be expected to have its own focus. More than simply borrowing validity types from psychometric, computer science, or qualitative research, there is an opportunity to address the unique nature of design science, wherein researchers develop artifacts to bring about desired outcomes and validate knowledge claims about those artifacts. Nevertheless, existing validity research in other disciplines has implications for validity in design science as it deals with general validity concepts and ideas. Our review of the validity literature led to the following findings that inform design science validity.

First, researchers agree that systematically evaluated and validated scientific research is necessary (Cohen et al. 2013). Even researchers who conclude that the label "validity" is not adequate for their research acknowledge the benefits of distilling and sharing successful and proven patterns of validation (e.g., Creswell and Miller 2000; Maxwell 1992).

Second, sciences seek to advance knowledge claims that constitute the contributions of the research (Collier-Reed and Ingerman 2013). A study can have many knowledge claims, with some being the focus of the study (primary claims) and others taken from prior research (secondary claims). Frequently, only the former is subject to validation in a study. In the validity context, therefore, a **knowledge claim** is an assertion about the phenomena of interest that captures the study's original contribution.

Third, when validating a knowledge claim, the object of the claim is commonly compared to a **reference entity**—an abstract or concrete object, the properties or behavior of which can be compared to the proposed idea or object to assess the quality of the latter. For example, in computer science, functional validity is achieved when "the model mimics the input-output behaviour of the real system to some acceptable level of accuracy" (Murray-Smith, 2015, p. 30), with the real system being the reference entity. Reference entities appear in various forms, including artifacts, existing natural objects (e.g., a human), and mental ideas. No matter the research claim, a reference entity always exists against which to validate the claim. Even for radical innovations or inventions, reference entities exist against which these can be compared.[2]

Fourth, to validate a knowledge claim, researchers engage in one or more evaluation procedures appropriate to that claim. An **evaluation (validation) procedure** is a set of tasks undertaken to provide evidence of the validity of the knowledge claim. Commonly, the procedure involves comparing on some dimension(s) the focal entity (e.g., the artifact developed in a design science study) to a reference entity, which may be material or abstract. These procedures are typically established by consensus within the discipline (Taylor 2013).

Fifth, validity is a matter of degree because the appropriateness and quality of the reference entity and the approach taken to validate it may vary. Thus, especially strong comparisons are conservative – performed against the best of knowledge, the *state-of-the-art* artifact or way of doing something, known as the criterion. Both quantitative and qualitative researchers (e.g., Moules et al. 2015; Taylor 2013) have argued that attaining perfect validity is impossible. Each research community determines the norms of what constitutes a sufficient outcome of comparison so that the corresponding knowledge claim can be accepted as scientifically valid.

[2] Consider great inventions of human history, including fire, the wheel, dynamite, the nail, and the printing press. All had manual or less innovative artifacts against which they could be evaluated. While the genius of many revolutionary inventions is the extent to which they depart from existing artifacts or ways of conducting tasks (the reference entity), when no existing reference artifact can be identified or it is impossible to access (e.g., it is proprietary), then other relevant artifacts or processes can be considered.

Finally, to systematize the agreed-upon and widely used patterns of reference entities, evaluation procedures, and norms for accepting evidence for specific types of knowledge claims, they are labeled and organized into **validity types**. Then, claiming a particular validity (e.g., convergent validity) can be used as a shortcut for suggesting that the corresponding claim (that two measures are, in fact, related) can be accepted as valid based on the extant scientific norms. For example, Cook and Campbell (1979) examined threats and presented approaches to ensure internal validity in social sciences, while Lincoln and Guba (1985) proposed steps for demonstrating the trustworthiness and authenticity of claims in qualitative studies. These validity templates suggest specific approaches for identifying the reference entities, comparing them, and presenting the findings. This makes it easier to reference and share validation practices, consistently use them, and improve them by identifying connections, overlaps, and gaps among validity types, thereby contributing to a cumulative tradition.

## THE DESIGN SCIENCE VALIDITY FRAMEWORK

Considering the foundations of validity, we now develop the Design Science Validity Framework.[3] We start with the knowledge claims about the focal artifact. Explicating these claims can support scholars in sharing and reusing useful validation procedures.

### Design Science Knowledge Claims

The nature of design science claims is rooted in design science as utilitarian scientific inquiry (Hevner et al. 2004). A key goal of design science is to create artifacts that address real-world challenges and to generate design knowledge related to these artifacts. For example, Hevner et al. (2004, p. 77) defined design science as research that "creates and evaluates IT artifacts intended to solve identified organizational problems." Since then, the understanding of design science has evolved beyond a pure organizational focus, as researchers increasingly seek to tackle broader societal and individual challenges (Weinhardt et al. 2020; Winter et al. 2014). Furthermore, many artifacts are sociotechnical, coupling software, hardware, and processes with individuals, groups, and organizations (Thomas et al. 2022). Thus, design science encompasses a wide range of social issues and corresponding innovative artifacts, an inclusive perspective that we adopt. What distinguishes design science from practice is the goal of developing knowledge related to building artifacts (Gregor and Jones 2007; Peffers et al. 2018). We therefore define **design science** as research that develops novel artifacts and relevant design knowledge to address individual, organizational, and societal challenges and opportunities.

Many types of artifacts are contributions in design science. Common artifacts are models, methods, instantiations, and design theories and theory components (e.g., constructs, design principles) (Gregor and Hevner 2013). Artifacts are complex entities with multiple interrelated parts or components.[4] They are also commonly taken to be components of larger systems, such as organizations or broader technical systems. Considering this diversity, we define a **design science artifact** as an abstract or concrete entity created as a focal contribution of a design science project to attain desired outcomes at the individual, organizational, or societal level.

The basic tenets that design science provides solutions to problems and generates prescriptive design knowledge are central to design science knowledge claims. Specifically, to assess the contribution of a given artifact, the artifact must be superior in some way (e.g., faster, more efficient, cheaper) relative to existing solutions (Padmanabhan et al. 2022), and the approach used to create the artifact must be specified (Hevner et al. 2004). To function as a science and allow others to benefit from the accrued design knowledge, researchers need to communicate knowledge about how the artifact was built—for example, through sharing code in a repository or describing how the artifact functions at a conceptual level (Burton-Jones et al. 2021; Hevner et al. 2024).

[3] In the following, we refer to the Design Science Validity Framework simply as the "validity framework" unless we intentionally refer to it by its full name.

[4] Each artifact type also has many subtypes. For example, the subtypes of *model* include frameworks, taxonomies, ontologies, simulations, and mathematical models.

Furthermore, the specific causal mechanisms that relate the design choices (parts of the artifact) to the desired outcomes can be provided to deepen the design knowledge arising from the development of the artifact (Gregor and Jones 2007; Peffers et al. 2018). Finally, as practitioners are unlikely to implement their solutions in contexts identical to the original research context presented in an article, it is valuable to specify when and under what conditions and boundaries an artifact is expected to attain its outcomes (Hevner et al. 2024). A key task for design scientists is to support other researchers and practitioners in reusing artifacts and design knowledge in new settings (e.g., Iivari et al. 2021).

The activities of conducting design science produce specific types of knowledge claims. A **design science knowledge claim** is a proposition about an artifact that asserts its contribution to science and society through a particular form or function. For example, Tiefenbeck et al. (2016) developed a shower meter and claimed that the particular indication of water and energy consumed (the form of the artifact) lowered energy consumption (the artifact's function) compared to showers lacking such indicators.[5] Consistent with the aims of design science, design science knowledge claims propose that an artifact built through a specified approach brings a desired outcome, that the outcome is caused by the specific design properties of the artifact, or that the outcome is expected to hold in a specific context or across different contexts. We label these types of knowledge claims *criterion*, *causal*, and *context*, respectively, and elaborate on them below.

**Criterion claims:** In every design science paper we reviewed, the authors make one or more claims about the *expected outcomes* of the artifact, such as its benefits or utility for addressing a challenge or opportunity. We define a *criterion claim* as a knowledge claim about the utility of an artifact. There is *always* an existing or alternative way of doing something, so criterion claims implicitly or explicitly present a comparison to such existing entities or processes (also known as the state of the art or criteria) (Padmanabhan et al. 2022). For example, authors may claim to have developed a more effective construct search engine (Li et al. 2020) or a novel design theory of tailorable technology (Germonprez et al. 2007). Such claims state or assume that, respectively, conventional search engines could work better by extracting theoretical constructs from papers or that typical design theories do not address tailorable technologies. We refer to these claims as criterion claims even though a claim itself does not need to explicitly specify the criterion.

Criterion claims are especially strong when the comparison artifact is the commonly agreed-upon state of the art (Padmanabhan et al. 2022). For example, Larsen and Bong (2016) compared the performance of their search algorithm to an EBSCO Host search engine criterion, whereas Li et al. (2020) made an arguably stronger claim by comparing their search algorithm to both EBSCO Host and Google Scholar—generally acknowledged as the state of the art for academic search. In our examination of design science articles (reported later), we found that *all papers* had criterion claims that implicitly or explicitly

[5] In the context of design theories, Gregor and Jones' (2006, p. 327) advanced a related notion of "testable propositions or hypotheses" explaining that these propositions about design theories are tested "through an instantiation, by constructing a system or implementing a method, or possibly in rare cases through deductive logic." Design knowledge claim is a broader concept: as we show in the paper, knowledge claims apply to any artifact, and to be validated, they do not require instantiating their components, even in the case of design theory.

indicated outcomes that were better than existing artifacts or processes.

**Causal claims:** In the long term, a criterion claim is insufficient to ensure scientific progress. After criterion claims have been established, researchers may (within a study that establishes a criterion claim or in subsequent studies) wish to deepen the design knowledge by substantiating claims about which design features of the artifact produce their claimed outcomes. We label such claims *causal claims*, given the fundamental interest of science in causes and effects (Salmon 1998). A causal claim is a knowledge claim about the extent to which specific parts of an artifact cause a specific utility.

While providing code or descriptions of an artifact helps support the value of an artifact and makes it more accessible to replication and extension, advancing causal claims can help build better artifacts and develop new theories of technology. For example, Abbasi and Chen (2008) advanced several causal claims when stating that CyberGate's feature sets were better at representing information types than baseline feature sets commonly used in prior systems (note the presence of a criterion claim as well). They conducted separate evaluations for each of the extended feature sets, namely topic, opinion, style, genre, and interaction information, to establish the causal influence of each on the artifact performance (the criterion coming from the performance of an artifact employing all the features).

Causal claims are not restricted to machine learning or even instantiated artifacts. In our evaluation, we evaluate causal claims about the value of parts of our own design science validities framework. In fact, given that most types of design science artifacts are not instantiated, qualitative methods play a major role in evaluating these artifacts. These methods, in turn, can be grounded in justificatory knowledge (e.g., kernel theory) that suggests the causal mechanisms connecting artifact features to target outcomes (Gregor and Jones 2007).

It is not always possible to know how or why something works, especially when dealing with complex technologies, innovative artifacts, and "inventions" (Gregor and Hevner 2013). Hence, not all papers make such claims; in our survey of design science papers, 50% made causal claims.

**Context claims:** Artifacts are always created in some context to address a given challenge or opportunity. However, the role of context is not always considered in design science. To support practitioners in implementing artifacts, *context claims* explicate the situations or conditions in which the proposed outcomes of the artifact are expected to hold. Accordingly, criterion and causal claims are generally considered more rigorous when evaluated in multiple contexts or contexts that are more like the intended use context of the artifact.

Not all design science papers make context claims. One exception is action design research, where the setting in which the artifact is implemented, a real organization (Sein et al. 2011), is the context in which any claims about the artifact are intended to apply. A context claim can be broadened (and hence, the scope of the criterion or causal claims strengthened) with suggestions that the artifact works beyond its original setting, such as in similar or even distinct settings. Further, a context claim can help delineate the generalization of findings. For example, Lukyanenko et al. (2019) explicitly claimed that their instance-based method of data collection is especially effective for open-ended tasks in large-scale citizen science projects, but also claimed that it is not expected to yield benefits for closed-ended collection in microtask crowdsourcing.

In our sample of design science papers, 14% of papers made context claims. This number may have been depressed by authors who did not explicitly make such claims, even when it was likely that such claims could have been supported with the evidence provided. If so, this represents an opportunity for researchers to consider advancing context claims (so that others reading their work can better appreciate the other contexts in which the solutions they developed might also be useful).

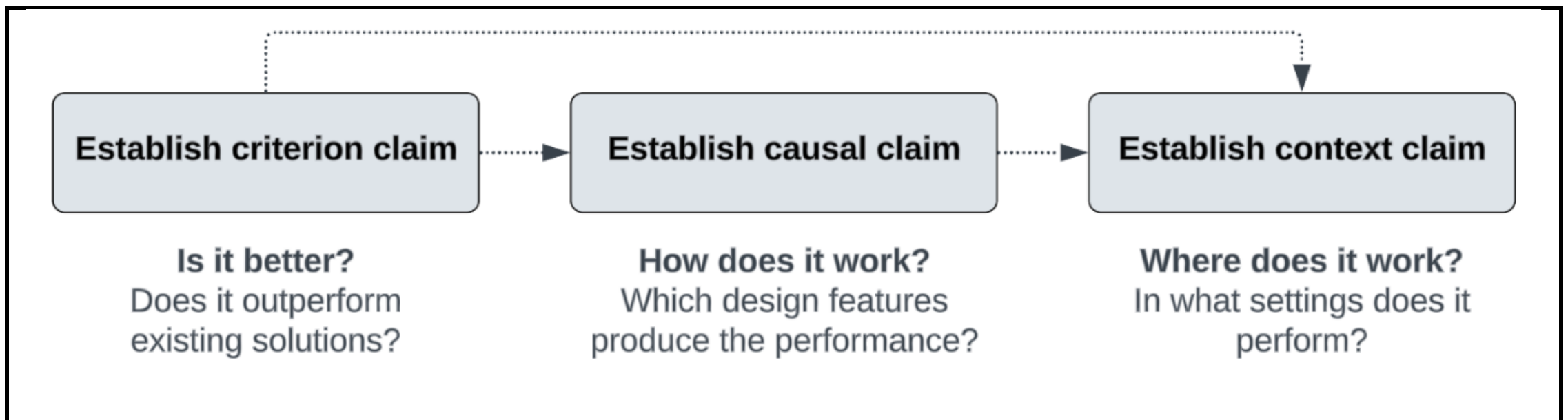

**Figure 1: Claims and Validation Process**

Figure 1 summarizes the validation process based on the importance and prevalence of knowledge claims. The figure implies that each paper makes at least one criterion claim. This may be sufficient for novel artifacts. As Gregor and Hevner (2013, p. 346) suggest, novel artifacts are "interesting applications where little current understanding of the problem context exists and where no effective artifacts are available as solutions." Then, depending on the artifact's novelty, a paper may additionally advance causal and/or context claims to improve the understanding of why and where the artifact works. Causal claims require the simultaneous or previous establishment of a criterion claim; context claims require either the establishment of a criterion claim or a causal claim.

The three types of knowledge claims can be made with respect to all types of artifacts. As discussed, abstract artifacts (e.g., design theory, conceptual models) can be instantiated, and (via instantiation) can be shown to have measurable impact on the world. Furthermore, qualitative techniques, such as counterfactual analysis and stakeholder interviews, can be used to evaluate criterion, causal, and context claims for abstract entities. The types of claims are driven by the desired knowledge contribution, not the artifact type.

## Design Science Artifact and Comparison Entities

We define **design science validity** as the extent to which knowledge claims about a focal artifact are supported by evidence. In line with the three types of knowledge claims, three general types of design science validity exist: criterion, causal, and context. By supporting their respective claims, these validity types advance the practice of design science in three important ways. First, *criterion validity* is used to support claims about the outcomes of designing and using the artifact, aimed at bringing about the desired change. Second, *causal validity* is used to support claims about the contribution of a specific design (artifact feature, part) to the utility of an artifact. Third, *context validity* addresses the extent to which knowledge claims hold for the intended context or additional contexts. Within these broad categories, we propose more specific validity types based, primarily, on the type of reference entity and, secondarily, on the nature of comparison between the focal artifact and the reference entity.

The focus of design science validity is on claims about a focal artifact. Some artifacts are *material* and others are *abstract* (Gregor and Hevner 2013, p. 341). For example, Tiefenbeck et al.'s (2016) shower meter is material whereas a theory, which is a system of concepts and propositions (Gregor and Jones 2007), is an abstract artifact.

Abstract artifacts can be used in developing material artifacts such as shower meters. Often, there are *two separate but related artifacts*—the theory (e.g., design propositions) and the material artifact generated from it (e.g., the shower meter). Artifacts are components of larger systems, such as *sociotechnical systems* of people and artifacts (Chatterjee et al. 2021; Winter et al. 2014). For many projects, it is important to consider the broader systems in which artifacts are embedded because the knowledge claims may be made with respect to systems-level impacts. If a researcher seeks to design the sociotechnical system itself as an artifact (see Thomas et al. 2022), criterion, causal, and context claims can be made about this sociotechnical system.[6] At the same time, if the focus is on the properties and behavior of the artifact (e.g., smart shower meter) embedded in a sociotechnical system (e.g., smart city), the system provides an implementation setting for the artifact, enabling context claims about the artifact to be made. Note that the focal artifact and the comparison reference entity exist within (typically different) sociotechnical systems, creating context validity challenges.

To derive the categories of reference entities, we considered the major types of artifacts produced by design science—constructs, models, methods, instantiations, and design theories, along with existing natural objects that can be used as criteria (vom Brocke, Winter, et al. 2020; Gregor and Hevner 2013; Hevner et al. 2004; March and Smith 1995). Table 3 outlines the list of reference entities to which a focal artifact can be compared. Having the reference entities allows us to explicate validity types and formalize the specific validation procedure the researchers can use to support their knowledge claims.

**Table 3. Categories of reference entities**

| Category | Subcategory | Details |
|---|---|---|
| **Instantiated entities**<br><br>Output-generating entities. When in use, they create change in the world, which can be measured and recorded. Outputs of reference entity may be in the past, present, or future. Other aspects than the outputs may also be compared. | **Criterion instance**<br><br>A real-world entity that becomes a standard against which a focal artifact may be compared. | Category includes criterion artifacts (e.g., the search engine Google Scholar), real-world objects (e.g., a human chess master), real-world processes (e.g., hiring employees), or sensors reporting on real-world states (e.g., temperature or rainfall).<br><br>Responsibility of the researcher is to justify that the criterion instance serves as a reasonable standard for comparison.<br><br>Instantiations may operationalize constructs, models, methods, and design theories into material solutions (e.g., apps, platforms, enterprise systems).<br><br>Criterion instances are used to demonstrate the superior performance of the focal artifact or the ability of the focal artifact to approximate the output, structure, or features of the criterion entity. |
| | **Manipulated artifact**<br><br>An artifact derived from the focal artifact; typically developed within the same study and manipulated to enable inferences from comparing two artifacts. | Constructed by removing or replacing a part of the focal artifact. |
| **Uninstantiated entities**<br><br>Objects of comparison that | **Theory** (and its components: constructs and design principles)<br><br>A system of concepts intended to explain, predict, or guide action. | *Design theories* and *their components* (e.g., constructs, design principles) and *non-design theories* (e.g., theories of explanation and prediction), which can provide design-relevant knowledge for focal artifacts. |

[6] Once a customer relationship management (CRM) system (the focal artifact) is placed in the broader social and physical system of an organization, we can evaluate attributes of the larger system, including the organization (e.g., customer satisfaction, fairness, sales, profit). Furthermore, by considering the organizations with CRM relative to those without CRM or the same organization before and after the introduction of the CRM, we can validate claims about the impact of the CRM or its components on outcomes of interest.

| | | |
|---|---|---|
| are blueprints for concrete material artifacts.<br><br>Unless instantiated, these entities do not produce material outputs. They commonly require interpretation by an agent such as a human expert. | **Model**<br><br>A conceptual representation of some aspect of reality created to increase understanding or facilitate action. | A model represents some aspect of reality; commonly ignores aspects not relevant to modeler; may introduce purposeful biases.<br><br>A model represents a domain where design interventions occur, or there are other design artifacts and is assessed with respect to utility or correspondence to truth. |
| | **Method**<br><br>A self-contained, logical sequence of steps used to accomplish a task. | Methods can be found in a variety of different areas. For example, requirements analysis, process control, and checklists. |

## Validity Types

**Criterion validity types:** As indicated in Figure 1, all design science projects should make a criterion claim to show that the designed artifact provides some benefit. Doing so requires engaging a criterion validity type. Criterion claims (and causal claims, as we show later) can be compared in two ways: through their efficacy or through their characteristics. Efficacy comparisons consider the similarity of the artifact's outputs to other output-generating entities, whereas characteristic comparisons assess the similarities among the characteristics of the artifact and its reference entities (Table 3). Therefore, criterion validity has two subtypes: criterion efficacy validity and criterion characteristic validity.

Criterion efficacy validity supports claims about the instantiated artifacts (Table 3) when the focal artifact and reference entity have comparable outputs. Criterion efficacy validity thus deals with criterion claims that are supported by comparing the efficacy of the focal artifact to that of an instantiated entity argued to represent a standard. These types of validity support claims that the outputs of the artifact (or the effects of such outputs on a sociotechnical system) have utility relative to the outputs or effects of a reference entity.

Two subtypes of criterion efficacy validity can be distinguished, depending on whether time is important. If so, predictive validity, with proper registration of the predictions before the results are generated by the reference entity, yields the strongest validity for claims. However, in design science, predictive validity is more often evaluated against future data that exist at the time of artifact creation but are not made available to the artifact until after the validation. If time is not a factor in the predictions, future data is not available, or a weaker validation is sufficient, concurrent validity may be evaluated. When a predictive claim is not made, concurrent validity may be employed to examine the outputs of the focal artifact relative to reference outputs. Often, available cases are split through cross-validation into 'train,' 'validation,' and 'test' sets, wherein the focal artifact's performance relative to true values in the 'validation' and 'test' sets are types of concurrent validity. In casual contexts, the term 'criterion efficacy validity' may be assumed to refer to concurrent validity.

Criterion characteristic validity deals with criterion claims that compare characteristics of the focal artifact to those of a reference entity argued to represent a standard. For characteristic validities, an agent (most often a human) is commonly involved in the evaluation because: the reference artifact does not produce outputs (as is the case for theories, models, and methods); the output is not directly comparable to the reference entity (as can be the case when comparing the outputs of two generative AIs); the artifact is instantiated but a characteristic of the artifact (such as its interface) contains the contribution; or the artifact is compared to the respondent's experiences with relevant artifacts or processes (such as with the evaluation of perceived usefulness). All such cases are ones in which a criterion characteristic validity is employed.

Characteristic validities allow the evaluation of aspects of the artifact that efficacy evaluations are unable to evaluate. They are more flexible and enable comparison of the non-identical outputs or outputs that cannot be automatically compared (e.g., perceptions of usefulness of a medical ontology or a website). Whether evaluation is based on the direct comparison of outputs or an artifact's characteristics, specific validity types are based on the comparison entity (Table 3). For example, criterion instance validity (instance validity for short) is the extent to which claims about an artifact relative to an instantiated reference entity are supported. Abbasi et al. (2018) engaged with criterion instance validity to validate a utility claim. The authors asked the organization to estimate the savings realized by implementing the system versus continuing with the current business process.

Similarly, theory validity addresses the extent to which claims about a focal artifact relative to a theory artifact are supported. For example, Lukyanenko et al. (2019) developed a citizen science platform with design features that they claimed corresponded to relevant principles from a design theory.

Criterion model validity (model validity) is the extent to which a focal artifact is consistent with a model. Although this validity type did not occur in the sample of design science papers we reviewed, the consideration of criterion model validity was present in our literature review. For example, Refsgaard et al. (2006, p. 1596) suggested that when developing the code for a simulation model (focal artifact), it is essential to establish that the "model code is in some sense a true representation of a conceptual model" (e.g., a model developed by experts) of the real-world system (e.g., of an ecosystem).

Finally, criterion method validity (method validity*)* compares a method to an existing method as a self-contained, logical sequence of steps used to accomplish a task. Criterion methods enable the evaluation of a focal artifact or part of a focal artifact against other entities (natural or artifacts, and their parts). Piramuthu and Doss (2017) provided an example when they evaluated their artifact—a protocol for the simultaneous authentication of multiple radio-frequency identification tags. They used formal proof to validate that the protocol met the strongest security requirement (Avoine et al. 2009).

**Causal validity types:** Like criterion validity, causal validity has two subtypes: causal efficacy validity and causal characteristic validity. Causal efficacy validity deals with causal claims supported through evaluation relative to the efficacy of a manipulated version of the artifact that has intentionally different parts (compared to the focal artifact). The manipulated artifact is typically developed in the context of the same study by the same research team and is manipulated to permit inferences arising from comparisons of the two artifacts. This is sometimes referred to as an ablation study, as introduced by Newell (1975), and commonly found in machine learning (e.g., Abbasi et al. 2012; Abbasi and Chen 2008; Etudo et al. 2017).

Causal characteristic validity addresses causal claims that are supported by comparison to the characteristics of a manipulated artifact that has intentionally different components. One way to achieve this difference is by removing or changing a component (a part). As with causal efficacy validity, the comparison entity is typically developed in the context of the same study by the same research team and is deliberately manipulated to enable inferences from comparing the two artifacts. For example, Umapathy et al. (2008) developed two versions of their focal artifact: one permitting the selection of integration patterns with the support of speech acts (focal artifact) and one with no support (reference artifact). The researchers performed an experiment in which participants were given a business process design task (with a list of specific task elements) to produce a modified Business Process Model and Notation (BPMN) model of an enterprise integration. The models were evaluated by experts and the perceived number of errors were compared.

Another approach relies on counterfactual reasoning (Collins et al. 2004). Comparing the focal artifact to a version that has a component removed can validate that the component does not have a causal impact on the performance of the artifact and can be removed for parsimony.

**Context validity types:** The context validity type supports claims about the setting and conditions (e.g., laboratory vs. real-world sociotechnical system, used by target users or their surrogates) in which the artifact is evaluated. The goal here is to assess the extent to which the criterion and causal claims related to the artifact are supported in a setting that is in some way like the target sociotechnical setting or generalizes across different settings. Hence, context validity evaluation procedures deal with the differences in conditions and settings. The context claims can either be made against the context of artifact evaluation relative to the target sociotechnical system or another context. This results in two subtypes of context validity: ecological (focusing on the attributes of the original evaluation context relative to those of the target sociotechnical system) and external (focusing on evaluation in additional sociotechnical systems).

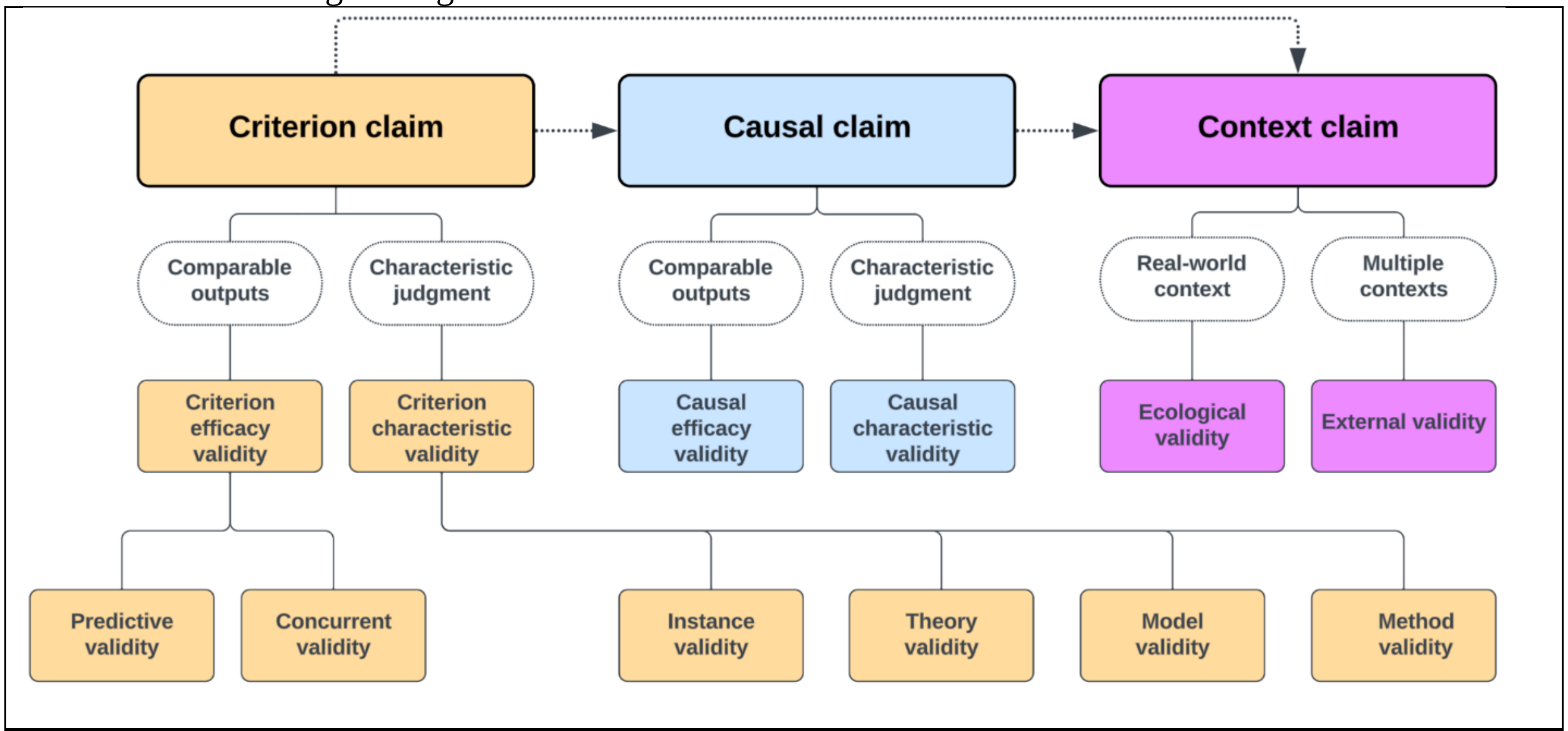

**Figure 2. Design Science Validity Framework**

A claim's ecological validity increases with the similarity between the evaluation sociotechnical system and the sociotechnical system for which the artifact is intended. Some design science approaches, such as action design research, tend to have high ecological validity because evaluations are performed in naturalistic organizational settings in which the artifact is being used. For example, Zaitsev and Mankinen (2022) worked with participants in Cambodia to develop an app that was used in a real-world setting.

External validity is defined as the extent to which criterion or causal knowledge claims are supported through an evaluation in more than one sociotechnical system. When the sociotechnical systems in which the focal artifacts are evaluated differ and the focal artifact retains its utility, higher external validity is realized. For example, Zaitsev and Mankinen (2022) used an action design research approach to develop an app to improve financial literacy through training. The original development occurred in Cambodia, but the artifact was later adapted and used in Nepal, providing external validity, with the authors concluding that "the original design, already flexible and minimalist, created according to the design principles, provided a good foundation for localization" (p. 106).

## The Design Science Validity Framework and Its Application

Figure 2 shows the Design Science Validity Framework. The framework establishes the relationship between the three types of claims about the focal artifact and the specific validity types used to provide evidence in support of these claim types, via specific types of comparison.

Table 4 formally defines each of the validity types.

| Table 4. Design Science Validity Definitions | |
|---|---|
| **Design science validity** | The extent to which knowledge claims about a focal artifact are supported by evidence. |
| **Criterion validity** | The extent to which knowledge claims about a focal artifact are supported through evaluation compared to a reference entity argued to represent a standard. |
| Criterion efficacy validity | The extent to which knowledge claims about the efficacy of a focal artifact are supported through evaluation compared to the efficacy of a material reference entity argued to represent a standard. |
| Predictive validity | The extent to which knowledge claims about the efficacy of a focal artifact are supported through evaluation of its accuracy compared to a reference efficacy that came into existence after the data used to create the artifact. |
| Concurrent validity | The extent to which knowledge claims about the efficacy of a focal artifact are supported through evaluation of its efficacy compared to a reference output that came into existence in the same period as the data used to create the artifact. |
| Criterion characteristic validity | The extent to which knowledge claims about the characteristics of a focal artifact are supported through evaluation of its characteristics compared to the abstract characteristics of a reference entity argued to represent a standard. |
| Instance validity | The extent to which knowledge claims about the characteristics of a focal artifact are supported through evaluation of its characteristics compared to the abstract characteristics of an instantiated reference entity argued to represent a standard. |
| Theory validity | The extent to which knowledge claims about the characteristics of a focal artifact are supported through evaluation of its characteristics compared to the characteristics of a theory reference entity argued to represent a standard. |
| Model validity | The extent to which knowledge claims about the characteristics of a focal artifact are supported through evaluation of its characteristics compared to the characteristics of a model reference entity argued to represent a standard. |
| Method validity | The extent to which knowledge claims about the characteristics of a focal artifact are supported through evaluation of its characteristics compared to the characteristics of a method reference entity argued to represent a standard. |
| **Causal validity** | The extent to which knowledge claims about the impact of a part of a focal artifact are supported through evaluation compared to a manipulated artifact missing that part. |
| Causal efficacy validity | The extent to which knowledge claims about the impact of a part of a focal artifact on that artifact's efficacy are supported through an evaluation of its efficacy compared to the efficacy of a manipulated artifact missing that part. |
| Causal characteristic validity | The extent to which knowledge claims about the characteristics of a part of a focal artifact are supported through an evaluation of its characteristics compared to the characteristics of a manipulated artifact missing that part. |
| **Context validity** | The extent to which criterion or causal knowledge claims are supported through an evaluation in a specific context. |
| External validity | The extent to which criterion or causal knowledge claims are supported through an evaluation in more than one sociotechnical system. |
| Ecological validity | The extent to which criterion or causal knowledge claims are supported through an evaluation in a sociotechnical system corresponding to real-world sociotechnical systems for which an artifact is intended. |

In general, different validity types are appropriate, depending upon the kinds of claims researchers wish to make. As our framework shows, there are three main types of claims, corresponding to the three main types of validities. Hence, our framework allows researchers to determine appropriate validity types based on the nature of the knowledge claims about an artifact. Once researchers have decided on a type of claim, they should follow that claim type from the top until they find an appropriate bottom-level validity type to support that claim.

For example, Koornneef et al. (2020) proposed a method to improve the process of identifying relevant information and options for resolving aircraft troubleshooting/maintenance issues between flights. The authors claimed that their prototype supported the faster retrieval of information relevant to issue resolution (a criterion claim, see Figure 2). They compared their prototype to the current practice in the industry of searching for relevant information in maintenance manuals. As the prototype and manual search methods provided comparable output (i.e., relevant issue information), this is an example of criterion efficacy validity with a specific type of concurrent validity (as time of comparison is assumed to play no role). To further bolster their claim, the authors performed an experiment in which advanced maintenance trainees were asked to locate relevant information using either the prototype or the manual method. The authors measured the search time in both cases and found that the prototype provided faster search times. If an evaluation supports the claim, only one leaf-node validity type may be necessary. Should the authors determine that their artifact is not novel enough to warrant publication in their target outlet, they may deepen the contribution by making a causal claim. While Koornneef et al. (2020) already had an efficacy validity evaluation, in a hypothetical scenario, they may then have added a causal efficacy validity by evaluating which parts of their artifact contributed most to the speed of retrieving relevant information.

**Table 5. Exemplar Validations**

| **Artifact and artifact type** | **Knowledge claim** | **Validity type** | **Evaluation** |
|---|---|---|---|
| Meth et al. (2015) propose a requirements mining tool, an *instantiation*. | Use of the tool allows users to identify requirements with greater accuracy than without the tool. | Concurrent validity | Experiment to evaluate whether users accurately identify all requirements using the tool. |
| Ramakrishnan et al. (2023) propose *design principles* for platform-enabled knowledge commons, a *theory.* | Principles have accessibility, importance, novelty and insightfulness, actability and guidance, and effectiveness. | Theory validity | Focus group evaluation of design principles. |
| Sedrakyan et al. (2017) propose a feedback-inclusive rapid prototyping (FIRP), a *method.* | FIRP simulation improves the understanding of the behavioral aspects of a model. | Model validity* | Factorial experimental design evaluating understandability of method. |
| Valecha et al. (2013) propose a shared vocabulary for message standardization in the emergency response domain, a *model.* | Model is complete and correct for covering the communication needs of emergency responders. | Model validity | Focus groups with emergency responder experts. |

* **Note:** The reference entity may be different from the artifact type, as it deals with establishing the baseline. For example, participants may draw upon their model of the world to assess some aspects of the proposed method.

This example illustrates how the framework represents existing design science practices, as well as how it can be used retrospectively to analyze an existing validation and better understand why certain choices were made. It can also suggest alternatives (e.g., evaluating prototypes by asking for human judgment, resulting in characteristic validity types), thus showing the possibility of using the framework to guide the design science validation process.

## EVALUATION AND APPLICATION OF THE FRAMEWORK

To assess the framework and simultaneously demonstrate how to use it, we applied the framework recursively to itself. A framework (a type of model) should have utility, which means, for example, that it must be seen as useful by its potential users (Hevner et al. 2004; March and Smith 1995). A framework should also be parsimonious while faithfully capturing the key domain's concepts (National Academies of Sciences 2022).

We formulate several knowledge claims about the validity framework, which we then validate. We advance several criterion claims, followed by context and causal claims to illustrate the usage of all types of claims and different types of validity. First, we claim framework utility as perceived by its potential users—design scientists. These are criterion and context claims supported by evidence of *model validity* and *ecological validity*. Second, we claim framework utility in systematizing existing *validation procedures* (named and unnamed), as supported by evidence of *model validity*. Third, we claim framework utility in systematically capturing and organizing existing design science validity types as proposed and defined across disciplines. These are criterion and context claims supported by evidence of *model validity* and *external validity*. Finally, we claim that the framework's components are sufficient for attaining utility, as supported by evidence of *causal characteristic validity*.

We further demonstrate the scope of the framework by considering sample work from the design science literature. We scope this demonstration by considering the artifact types proposed by March and Smith (1995) and Gregor and Hevner (2013): instantiations, design theory, methods, and models. For each of these artifact types, Table 5 presents a sample paper and identifies the artifact, claims, and validity presented by the authors.

### Claim 1: The criterion claim and context claims with target users

For Claim 1, we evaluated criterion and context claims by undertaking two applicability checks (Rosemann and Vessey 2008), a common type of design science evaluation (e.g., Li et al. 2020; Lukyanenko et al. 2019). The applicability check is a composite evaluation touching on several validity types. However, for evaluating the validity framework the applicability check is effective for addressing *model validity*, wherein participants compare our framework to their model of an ideal framework and other relevant frameworks. Because it engages the likely users of the framework and asks them to employ it in the context of their work, the applicability check also provides evidence of *ecological validity*. Our evaluation took place in the context of design scientists who had extensive experience with evaluation and exposure to existing evaluation frameworks.

Prior evaluation frameworks, including those of Prat et al. (2015) and Venable et al. (2016), focus on patterns of evaluation and on contemporaneous and post hoc strategies for evaluation, respectively. By coding six dimensions of evaluation techniques and criteria, Prat et al. identify specific techniques (e.g., observation, description, experiment, dynamic analysis) and specific research approaches (e.g., qualitative, experiment, simulation, metrics) that are used in evaluation. Venable et al.'s (2016) framework focuses on the artificial/naturalistic context in which evaluation takes place and how those evaluations can be formative (taking place during the project) or summative (taking place at the conclusion of the project).

In contrast, our validity framework provides the underlying scientific reasoning for validating knowledge claims through the evaluation process. By this, we mean the degree to which evidence gained through an evaluation process supports the knowledge claims made by researchers about the focal artifact. Furthermore, our analysis shows that evaluations in design science are always comparisons: performance criteria compared with existing criterion artifacts (or existing ways of accomplishing the same outcome), desired artifact characteristics compared with existing artifacts, or comparisons within and/or between contexts of artifact application. In articulating the relationships among criterion, causal, and context validities, our validity framework goes beyond prior evaluation frameworks. We show the processes by which evaluations are performed and demonstrate how the validity of types of knowledge claims is established. By making these relational comparisons explicit, our framework: provides meta-categories (e.g., criterion, causal, and context) for evaluations; organizes existing validities into these meta-categories; and identifies the specific comparisons and the *claim—evaluation—evidence* argumentation structure of the evaluations themselves.

In the first applicability check, the validity framework was presented at a regional information systems workshop attended by both behavioral researchers and design scientists; our focus was on the initial assessment of utility and making any adjustments before we engaged the final cohort of design science experts. Fifteen participants were trained in using the validity framework and tasked with organizing four validity definitions and four design science evaluations into the framework and responding to an open-ended survey. The results suggested that the framework would be useful to design scientists and other researchers. For example, the framework was described as "a thought-provoking and refreshing perspective." Other comments noted that the framework was quite complex, especially for researchers with limited design science experience. In response, we added a process for selecting which validity types to use in a research project and streamlined the framework.

After incorporating the feedback into the framework (the final version of which is shown in Figure 2), we performed the second applicability check among leading design scientists to evaluate the framework's importance (criterion claim), accessibility (context claim), and suitability (context claim).

The 11 participants were design scientists with varying levels of post-Ph.D. academic experience (an average of 13 years). For those with Google Scholar accounts (nine participants), their average citation count was 4,981. An examination of their papers found that 36% had cited Prat et al. (2015) or Venable et al. (2016). The applicability check was conducted online to include scholars from various locations, research areas, and sub-communities of design science. These participants employed a variety of methods, considering diverse research topics and types of analysis. Their research covered a broad set of topics, including data and knowledge modeling, analytic and machine learning modeling, data science, business process change, simulation, and design theory. We specifically invited participants across a broad spectrum to include participants across design science as a whole. Participants were regular contributors to WITS, DESRIST, and other design science conferences. The stages of the applicability check are described below.

**Preparation:** Participants were asked to describe examples of the validation processes they used in ongoing research or a recently completed project. These existing processes, part of the mental models of the experts, were the benchmarks against which participants assessed our framework.

Overall, participants exhibited a consistent understanding of the need for artifact evaluation and validation but demonstrated a limited understanding of the types of validity that aligned with their evaluations and lacked a vocabulary to describe their validations. In multiple cases, validity types were discussed in psychometric terms (e.g., construct validity) or terms related to utility and efficacy (e.g., proof of concept, proof of value). Most participants focused on the processes by which evaluations were performed (e.g., field experiments, lab experiments), rather than the role of validations in supporting claims about their artifact.

**Introduction to the Design Science Validity Framework:** Participants then attended an hour-long online session during which the author team presented the validity framework.[7] Following the explanation of the framework and the extended example, participants were asked to use our process to determine validity types in their projects.

**Feedback:** When finished the task, the participants filled out a survey containing six open-ended questions and a set of questions about artifact usefulness and their intention to use, applying Likert-type scales from Venkatesh et al. (2003). The responses, both qualitative and quantitative, indicated that participants considered the validity framework to be important, accessible, and suitable (Rosemann and Vessey 2008) for meeting the identified needs of the community and believed it would be useful for their own research.

Rosemann and Vessey (2008) define importance as research "that meets the needs of practice by addressing a real-world problem … in such a way that it can act as the starting point for providing a solution" (p. 3). We consider the process of evaluating this aspect to be an instance of criterion validity. In our context, design scientists are practitioners in the real world who themselves need to validate knowledge claims about artifacts. Accessibility is a criterion validity that "encompasses whether the research is understandable, readable and focuses on results rather than the research process" (p. 3). Finally, suitability is defined as the extent to which the research can "[meet] the needs of practice" (p. 3), which we take to mean the extent to which researchers view the framework as appropriate for the target context. These three evaluations all address model validities as well as context validity, given their evaluation in a setting similar to the real world.

Participants agreed that the framework is important for clarifying and providing structure to the increasingly complex requirements for validation methods, criteria, and strategies. This importance was evidenced by comments such as "it supports a systematic approach to validation" and "a framework like this can give researchers an accepted standard to point to as they try to validate their own design science artifacts." Other comments acknowledged that the framework reduces complexity, connects knowledge claims to validation strategies, and provides details useful for elaborating validity claims. All these comments support our criterion claim.

Despite only having a brief introduction to the framework, accessibility was evidenced by comments such as: "It is a good thing I have learned about your validity framework early in my project." One participant found the framework to be immediately accessible, stating: "I can use the framework …. the framework provided a validation that my evaluation is good enough, but I should have elaborated a few more details in my evaluation for reviewers."

The final applicability precursor, suitability, was also reflected in comments: "As a DSR scholar and author, the framework helps me to plan the validation of my work early on. … As a reviewer and editor, I will be able to point to a shared understanding of validity in DSR. When assessing the value of a DSR paper, I can use the framework to identify strengths and potentials for improvement. Or to prompt the authors to share the aspects of their DSR work which they have excluded from the paper but are relevant to establish its validity."

[7] The presentation is available in the transparency materials at https://osf.io/ca6vg/.

Two concerns were raised regarding the future implications of the framework. The first was that the framework could be used to inhibit the publication of design science because “reviewers may disagree on which validity types are required.” Participants agreed that this is not a flaw in the framework itself. It is inappropriate to use the framework as an argument for additional evaluations. As noted above, additional evaluations are justified only when either the evaluations performed are not aligned with the knowledge claims made or the contributions of the paper with the existing knowledge claims are deemed by the review team to be insufficient. In the latter case, any additional evaluations reviewers propose should be appropriate to additional knowledge claims.

The second concern was the potential for increased costs and time requirements for producing and reviewing design science work because of increasing demands for validity procedures. We updated our initial framework to better explain the knowledge claims and the process of applying the framework to protect against such use of the framework.

Participants responded on a 7-point Likert scale about the usefulness of the framework ($\mu = 5.64$; $\sigma = 1.04$), a common applicability check question in design science (e.g., Li et al. 2020; Lukyanenko et al. 2019). They were also surveyed about their intention to use the framework ($\mu = 6.15$; $\sigma = .92$) relative to their existing process, another criterion claim that, when evaluated in the target setting, doubles as a context claim. For usefulness, one outlier indicator reflected the second concern about whether the framework would (at least initially) slow down research tasks. Without this indicator, usefulness increased markedly (from $\mu = 5.64$ to $\mu = 6.14$; $\sigma = 1.03$).

The one participant who worried that the framework might be difficult to use also intended to use it, stating that “the existence of a framework like this can give researchers an accepted standard to point to as they try to validate their own design science artifact [claims].” All participants indicated that they intended to use the framework once it became available. We therefore concluded that the applicability checks established *model validity* and added initial evidence of *ecological validity*.

## Claim 2: The criterion claim of framework completeness

We validated the claim that the validity framework is complete in representing existing evaluations in published design science articles. Since the framework is a model, it should be able to represent these evaluations yielding *model validity*. We consulted two sources to identify a population of relevant papers against which to sample. First, we used the 121 design science papers identified by Prat et al. (2015) from the Association for Information Systems (AIS) Senior Scholars’ Basket of Eight journals (April 2004 to December 2013). Second, we analyzed 1,233 additional candidate papers from the AIS Basket of Eight from 2014 to 2017. We then manually coded these 1,233 papers based on the inclusion schema of Prat et al. (2015). Eighty-six papers were coded by the third author as design science and cross-checked by the second author. There was 100% agreement that all were design science papers. The original 121 papers from Prat et al. (2015), plus our 86 papers, yielded a total of 207 papers published between April 2004 and December 2017.

To capture articles that followed the design science approach but did not explicitly use the phrase “design science,” we created feature sets based on a list of design science keywords as well as citations of top design science papers. We then trained a machine learning model for distinguishing relevant manuscripts in the full set of manuscripts, based on the process described by Larsen et al. (2019). We applied a combination of machine learning and manual evaluation to expand the years of coverage from 1994 to 2019 and the sources to include the AIS Basket of Eight, as well as *Decision Support Systems*, *ACM Transactions on Information Systems*, and the proceedings of the International Conference on Design Science Research in Information Systems and Technology (DESRIST) and the International Conference on Information Systems

**Table 6. Validity Definitions (shaded cells indicate "not applicable")**

| Name | Part | Output | Artifact | STS |
|---|---|---|---|---|
| Design science validity | | | | |
| Criterion validity | | | | |
| Criterion efficacy validity | | | 19 | |
| Predictive validity | | | 3 | |
| Concurrent validity | | | 4 | |
| Criterion characteristic validity | | | | |
| Instance validity | 0 | 0 | 5 | |
| Theory validity | 0 | 0 | 0 | |
| Model validity | 0 | 0 | 0 | |
| Method validity | 0 | 0 | 10 | |
| Causal validity | | | | |
| Causal efficacy validity | 24 | | | |
| Causal characteristic validity | 7 | 0 | 1 | |
| Context validity | | | | |
| External validity | | | | 7 |
| Ecological validity | | | | 1 |

(ICIS). After downsampling the last four sources, we ended up with 527 design science papers.

From our corpus, we randomly sampled 32 empirical articles, which the fourth and sixth authors independently coded, identifying 79 evaluation descriptions. The first and the third authors then independently coded each evaluation description using the framework. The coders agreed in 76.3% of cases, resulting in a Cohen's kappa of .703. Disagreements generally concerned what the original authors intended to claim rather than how to interpret the validity framework. From this exercise, we concluded that evaluations found in the papers reviewed could be classified using the validity types in our framework, demonstrating *model validity*. However, two validity types, *theory validity* and *model validity*, were not used in the papers we sampled (Table 6).

## Claim 3: The criterion and context claims of representational power

The formalization of shared disciplinary conceptualizations accelerates scientific progress (National Academies of Sciences 2022). Other disciplines also develop and validate artifacts, and the validity framework must be capable of representing the validities used to evaluate these artifacts.

Claim 3 is that the validity framework is complete in representing design science validity types, not only for information systems, but also for behavioral science, engineering, and medicine—when these disciplines are validating claims about their artifacts. By evaluating the criterion claim that the framework is complete in its ability to represent design science validity definitions, we establish *model validity*. By showing that our framework applies to other disciplines, we establish *external validity*.

The first task was to identify the specific design science validity types already proposed in the literature. No sizeable existing set of general validity types was found. We thus built such a dataset for further refinement of a subset of validity types. We started by identifying and collecting validity types and definitions from various fields, including social sciences, engineering and computer science, and medicine. This was compiled over a three-year period by the first author and a team of research assistants. The *initial sources* were documents containing sets of validity types, such as the standards for educational and psychological testing (e.g., American Educational Research Association et al. 2014). The largest collection of validity names, 168, was provided by Newton and Shaw (2014). The vast majority of these were in a large table as an exhibit of the intractable nature of validity types. Still, many did not appear in any available database and could not be found using a search engine.

The second task was to obtain definitions for the identified validity types. We relied primarily on scientific books and articles. For every source, the section containing the validity was further examined to identify more *candidate validity types*. A candidate validity was a concept that the author either stated to be a validity or a concept not stated as such but listed with validity types (e.g., *mundane realism* as closely related to *ecological validity*). We did not question the authors' statements. In each case, we recorded new candidate validity types. In total, 2,418 candidate validity types emerged from approximately 7,500 manually examined sources. They were categorized by name before each category set was examined to find homonyms (same name but different meaning), yielding 418 distinct validity types. This literature review was used to generate the first version of the Design Science Validity Framework.

In the third task, we examined all articles published in the AIS Basket of Eight journals (Lowry et al. 2013) from 1994 through 2017. A total of 6,083 articles were analyzed by applying 216 regular expression queries representing validity, reliability, and related concepts such as generalizability, which yielded 73,365 sentences. Sentences were ordered by the number of hits, with 9,707 sentences containing more than two hits on the regular expressions manually examined by the first author. This analysis only yielded 23 additional validity types, for a total of 441 candidate validity types, suggesting that the original process had been thorough. We removed from further evaluation any candidate validity type not in common use for which we failed to locate five definitions from different sources. One hundred and fifty-eight candidate validity types were removed in this step.

During the fourth task, the first and second authors independently read the five definitions for each validity type and selected one or two definitions that represented the overall aspects expressed by the other definitions. The decisions were discussed (and recorded) and disagreements were resolved to obtain agreement on one or two definitions to represent the validity type. Because there were many cases where highly similar or even identical validity types existed, calculating interrater reliability metrics was not appropriate. Eleven candidate validity types were eliminated in this step because they did not fit our definition of a research validity (for example, law-based validities) or because a clear definition had not emerged.[8]

For the fifth task, with a final set of validity types specified, the first and second authors followed our established definition of design science validity and independently coded all validity types as "design science validity type" or "other validity type." The coders reached a 90.4% agreement with a Cohen's kappa of .79. The coders discussed and resolved any disagreements. Almost all disagreements stemmed from the efficacy validity types, where one coder employed a more inclusive interpretation. In total, 79 design science validity types were found.

From the design science validity types, we removed non-leaf validity types, which combined multiple other validity types. After removing these validity types, 70 definitions remained for categorization in our framework. Of these, 23 (33%) came from behavioral science outlets, 37 (53%) from engineering and computer science, and 10 (14%) from medicine; 22 (31%) were drawn from the sample of IS journals.

The first and third authors independently coded all validity definitions into the framework and agreed on 79.2% of cases (Cohen's kappa = .731), which included cases of partial agreement coded as non-agreement (for example, criterion efficacy validity vs. predictive validity). Disagreements were discussed and resolved. All definitions fit into a validity category of the framework (left column of Table 7).[9]

[8] Justification for each removal is available in transparency materials.

[9] Note: validity metrics were included in the analysis but, because they do not employ a claim, they are not themselves validities. We therefore excluded the following metrics from classification: accuracy, area under the curve, completeness,

| **Table 7. Validity Definition Coding** | |
|---|---|
| **Design science validity** | **Validities in the literature** |
| Criterion validity | |
| Criterion efficacy validity | Absolute validity, criterion group validity, criterion validity, criterion-oriented validity, criterion-related validity, decision validity, diagnostic validity, discriminative validity, empirical validity, lower-order validity, operational validity, pragmatic validity, procedural validity, postdictive validity, replicative validity, retrospective validity, application validity |
| Predictive validity | Predictive criterion validity, predictive validity (2)*, predictive criterion-related validity |
| Concurrent validity | concurrent criterion validity, concurrent criterion-related validity, concurrent validity, cross-sectional validity, relative validity |
| Criterion characteristic validity | |
| Instance validity | Observational validity, physical validity |
| Theory validity | Aetiological validity, theoretical validity, instantiation validity |
| Model validity | Conceptual model validity, functional validity, structural validity, semantic validity |
| Method validity | Algorithmic validity, consistency |
| Causal validity | |
| Causal efficacy validity | |
| Causal characteristic validity | |
| Context validity | |
| External validity | Pragmatical validity |
| Ecological validity | Behavioral validity, ecological validity |

**Note:** *two slightly different versions of predictive validity referencing the same validity type.

Thus, the evaluation yielded strong evidence for the validity of Claim 3 in terms of both the criterion claim and the context claim, strengthening *model validity*, and providing *external validity*. It did not support the need for correct rejection, detection rate, F1-score, fall-out, false alarm, false discovery proportion, false negative, false negative rate, false omission rate, false positive, false positive rate, hit, hit rate, Matthews correlation coefficient, miss, miss rate, negative predictive value, positive predictive value, precision, recall, sensitivity, specificity, true negative, true negative rate, true positive, true positive rate.

the two causal validities included in the framework. However, our evaluation in Section 5.2 has established the need for the causal validity types. The lack of attention to causal validity in the existing validity type definitions suggests an opportunity to contribute to design science validity and, in turn, evaluation. In this sense, it supports the need for templates in design science, as proposed by Peffers et al. (2008).

## Claim 4: The causal claim that the framework is parsimonious

The Design Science Validity Framework was developed iteratively, as is common in design science. As we wanted to remain inclusive, the initial version of the framework was more extensive than the one reported in this article. For example, the initial framework contained a validity type termed "requirement validity"—a type of criterion characteristic validity employed when comparing an artifact to a requirements document or a user's expectations and experience. However, requirements are not design science artifacts that capture contributions to science and society, and we were unable to clearly classify requirements validity as a distinct validity type. We therefore engaged in a *causal characteristic validity* evaluation to examine the need for this validity type (supporting the claim that our framework without this validity type is parsimonious and has no more components than needed).

Requirement validity was introduced to address knowledge claims against explicit requirements (e.g., function, ease of use, form) and implicit requirements (e.g., needs, goals, or experiences with a similar class of artifacts). *Requirement* is a common artifact produced in software development. Still, it does not feature on common lists of design science artifact types (e.g., Gregor and Hevner 2013; March and Smith 1995). However, we initially considered it to be a major reference entity of our framework, erring on the side of being more inclusive and conservative. As our understanding of the validity domain expanded over time, we began to question whether *requirement* was needed. We choose to resolve this issue by making a formal claim that *requirement* is a necessary validity type for the framework to be complete.

To validate this claim, we chose an evaluation approach based on counterfactual reasoning. The first and the third authors cooperatively coded two validity definitions previously coded as "requirement validity" in the original framework. First, we examined "application validity," defined as whether a simulation model corresponds to its purpose and requirements; i.e., the likelihood that the model produces outputs that reflect some external artifact or sociotechnical system. Based on our improved understanding of the framework, this validity instance was recoded as a *criterion efficacy validity*. The second existing definition was "semantic validity," defined in part based on the appropriateness of the category definitions. We realized that the reference entity was the semantic evaluators' understanding of an equivalent model or sociotechnical system category, suggesting that this was a *model validity*. The same logic became clearer when examining seven design science evaluations initially categorized as "requirement validities." In most cases, the initial evaluation focused on terms such as "requirement" or "satisfaction" and "acceptability." Often, the evaluations were poorly described by the original authors and unclear in terms of the actual reference artifact or sociotechnical system. In this causal validity evaluation, we focused on what we believed the researchers considered to be their reference artifact. For the seven evaluations, we found five cases of *model validity* and two cases of *instance validity*.

Thus, this evaluation showed that the validity framework could be made more parsimonious without losing representational capability. As such, the counterfactual evaluation for *causal validity* demonstrated that the removed parts of the framework were not causally implicated in the performance of the framework. The other validity types were all built around the commonly acknowledged artifact types and were necessary to classify the evidence, suggesting *causal characteristic validity* for the original claim of parsimony for the remaining parts of the framework. Therefore, it is unlikely that additional causal validity evaluations would enable the removal of additional parts. As a result, no further iterations were deemed necessary.

## DISCUSSION AND IMPLICATIONS

This research contributes to validity in design science that could be expanded to other applications. There are both theoretical and practical implications of our work.

### General contributions

A critical part of design science is validating knowledge claims about the focal artifacts. This research articulates why validation procedures are needed and provides a process by which they can be identified and enacted through various validity types.

We propose the Design Science Validity Framework. It maps validity types to characteristics of knowledge claims, aiding authors in formulating and communicating their knowledge claims and the evidence supporting them. Based on the largest-ever review of validity literature, the Design Science Validity Framework is a comprehensive framework for validating knowledge claims about artifacts. The framework provides a standard vocabulary for research validity reporting. It has external validity in that it successfully represents all identified artifact-related validity evidence from information systems, behavioral science, engineering, computer science, and medicine.

The framework guides researchers in identifying the knowledge claims about artifacts by considering the branches of the hierarchy of validity types, thus strengthening the rigor and contribution of design science projects. By providing a process and nomenclature for validating knowledge claims, the framework can be applied to any type of artifact creation and evaluation. Explicitly connecting validity types, evaluation processes, and supporting evidence to knowledge claims should be useful for researchers and reviewers in design science, as well as for those adopting other research approaches. Hence, our framework can help forge ties between design science and other types of research. The explicated knowledge claims can also be used to extend contributions in prior work. This could be done, for example, by making causal and context claims about the artifacts or adding additional criterion claims with more recent comparison entities.

### Findings and implications

Constructing and evaluating the Design Science Validity Framework led to several notable insights that demonstrate its value. First, design scientists have historically used ad hoc evaluations to support implicit claims about validity of the artifact itself. To center and broaden validity as an aspect of rigor in design science, we shift these evaluations to establishing the validity of explicit knowledge claims about the artifact. These claims will have varying degrees of supporting evidence, the sufficiency of which will be established by the community and potentially contested and changed over time.

Second, there is a surprising lack of comprehensive discussion of design science validity, even though the evaluations and types of validity are well-established and understood in other disciplines. Published design science uses a narrow range of validity concepts, largely focusing on efficacy measures and characteristic validity types, suggesting that validity has been underutilized. In our review of the literature, the most frequently occurring terms were accuracy, precision, recall, specificity, true positive, and false positive (concepts related to the confusion matrix). These are, in fact, metrics rather than validity types, but are used in establishing efficacy validity. In the sample of publications analyzed, these validity metrics were most frequently related to the evaluation of machine learning models, which, of course, does not represent the scope of design science.

Third, some validity types used in design science have been adopted from other disciplines. However, this does not facilitate a holistic evaluative approach for design science. For example, validity in psychometrics and econometrics are strongly focused on measurement, and some types are useful in design science (e.g., internal validity to support causality claims in experimentation). However, psychometric validities are insufficient to support the range of knowledge claims in a design science project.

Finally, as confirmed in our applicability check, a major challenge for researchers who create artifacts lies in the uncertain nature of validation and the tendency to perform validation activities implicitly with little sense of the underlying structure of such validations. The validity framework provides researchers with a structured template and a set of carefully explained validity types with explicit, standardized definitions. The validation process allows researchers to gather evidence that supports a validity type explicitly connected to a knowledge claim about a designed artifact.

## Practical suggestions

Summarizing the arguments and findings of our paper, we make the following recommendations for researchers who are interested in developing artifacts to contribute to science and society.

**State knowledge claims explicitly**. To attain high levels of practical utility and scientific replicability, researchers should explicate knowledge claims about their artifacts. This can help frame research contributions and guide validation, because validation depends on articulated knowledge claims. In addition, explicit knowledge claims support the accumulation of knowledge.

Given that design knowledge can evolve through iterative refinement of an artifact (Tuunanen et al. 2024), claims can emerge at various stages during iterations. For example, some claims can be made prior to the development of an artifact, while other claims can be made after deploying an artifact in some context and observing outcomes. However, the latter claims should not be considered validated until they have undergone an appropriate validation procedure, possibly in a subsequent iteration of the artifact. What is important is that a knowledge claim is formally evaluated independent of the process that generated the claim, which requires that claims be stated explicitly.

**Make claims commensurate with the intended contribution.** The Design Science Validity Framework should not be used to justify excessive validations. It is often impossible/unnecessary to state every claim about the artifact and not all claims can be subjected to validation (e.g., due to the difficulty in performing comparisons, or acquiring suitable comparison entities). The question of how much evidence is necessary depends on the context and specific characteristics of a problem.

Researchers should make at least one criterion claim about the artifact, striving to make comparisons against state-of-the-art artifacts or processes. This may be sufficient if the artifact is particularly novel, such that little is understood about what makes it effective or about additional (beyond the original) settings where it can be deployed. Beyond this, causal and context claims strengthen the research contribution because they deepen design knowledge and help practitioners reliably and safely apply the research findings in diverse settings.

**Ensure every knowledge claim is validated**. If an original claim is formulated about the artifact, it should be validated. For example, if some component of an artifact is claimed to cause a specific outcome, then establishing causal validity is appropriate. While a single validation does not prove a knowledge claim, validating claims increases the likelihood of producing reliable design knowledge.

As for how much evaluation is needed to establish validity of a claim, Galison and D'Agostino (1987) present a convincing argument that the sufficiency of evidence is a matter of community agreement. The number of validation activities required is commonly not based on a specific rule but on a consensus regarding the assembly of "persuasive arguments, ones that will 'stand up in court'" (Galison and D'Agostino 1987, p. 227). Researchers and review teams reach such consensus during the review process. What is important is that the validation procedures undertaken are appropriate for validating the knowledge claims made in a paper.

**Validate claims throughout the artifact journey**. Many artifacts emerge through iterative processes via experimentation, tinkering, or gradual improvement. Some projects involve multiple stages and produce different artifacts during these stages (e.g., conceptual model, system prototype, beta version, final system in production) (Tuunanen et al. 2024). Knowledge claims about the resulting artifacts can be made throughout this process. Formative validation of these claims can help in further artifact refinement and improvement. The "intermediate" validations can be informal, such as using a convenience sample of prospective users or an easy-to-obtain criterion. However, to ensure that the resulting artifact reliably contributes to science and practice, the claims about the *final* or *public* version of the artifact should be subject to rigorous summative validation.

**Ensure appropriate validity types are used.** With the establishment of the Design Science Validity Framework, a researcher can refer to the framework during validation of their knowledge claims. The framework organizes diverse validity practices into a coherent reference system. It shows what validity types are appropriate for each claim type and suggests the comparison entities and comparison procedures reasonable for these validity types. As the community continues to apply and refine these validity types, their robustness is expected to grow, giving researchers a stronger foundation upon which to build their research.

The Design Science Validity Framework provides opportunities for future research. First, the framework is extensible, meaning it can accommodate additional validity sub-types that might be proposed by the research community (e.g., further refinement of model validity based on types of models). Second, it is possible to better track patterns of validation in design science, and identify gaps and opportunities (e.g., the need for more context claims). A related possibility is improving validation procedures by ensuring that appropriate validation practices are systematically captured for their respective validity types. Finally, researchers can investigate the applications of the framework in design science projects and report results related to the usefulness and long-term impact of using the framework on the maturity of design science and its integration with other research traditions.

## CONCLUSION

This research defines validity for design science and proposes the Design Science Validity Framework and a process for its use. The framework, derived from an extensive review and analysis of the literature on validity, identifies and organizes implicit and explicit validity types into three main categories: *criterion, causal,* and *context.* The validity categories are intended to assist researchers in rigorously obtaining and presenting evidence of their knowledge claims. Mapping knowledge claims onto the validity framework should support scholars and help to connect the scientific knowledge related to information systems artifacts. The framework itself was validated by evaluating its own knowledge claims and providing evidence for the relevant validity types that support those claims.

## Acknowledgements

The authors wish to thank the Senior Editor, Balaji Padmanabhan, the Associate Editor, Stefan Seidel, and the anonymous reviewers for their work on prior versions of our paper. We appreciate the feedback received from participants at workshops at HICSS and ECIS as well as the University of Colorado, the University of Arizona, Arizona State University, and the University of Sydney.

**Funding.** This research was supported in part by funding from the U.S. National Institutes of Health to Kai R. Larsen [Grant 3U24AG052175-08S1] and by funding from the Natural Sciences and Engineering Council of Canada to Jeffrey Parsons [Grant RGPIN-2020-04916].

## BIOGRAPHIES

**Kai R. Larsen** [0000-0002-8812-9866] is a Professor of Information Systems in the Leeds School of Business, University of Colorado Boulder. He is a courtesy faculty member in the Department of Information Science, a professor at the University of Agder, and a Research Advisor to Gallup. Kai is most known for providing a practical solution to Edward Thorndike's (1904) Jingle Fallacy and for his contributions to the Semantic Theory of Survey Response (STSR). He received the INFORMS *Design Science Award* for 2019, the *Herbert A. Simon Design Science Award* for 2020, and the AIS *Education Innovation Award* for 2023.

**Roman Lukyanenko** [0000-0001-8125-5918] is an associate professor at the McIntire School of Commerce, University of Virginia. His research interests include data management and research methods (validity and artificial intelligence in literature reviews). Roman actively develops ideas, tools, and methods to improve data management and research practices. These solutions received major awards, including INFORMS Design Science Award, Governor General of Canada Gold Medal, Hebert A. Simon Design Science Award. Roman's research appeared in Nature, MIS Quarterly, Information Systems Research, ACM Computing Surveys. His 2019 paper on quality of crowdsourced data received the Best Paper Award at MIS Quarterly.

**Roland M. Mueller** [0000-0002-8706-7763] is a Professor of Information Systems at the Berlin School of Economics and Law in Germany. He earned his Ph.D. in Information Systems from the Free University of Berlin. His research interests include ontologies and large language models for literature reviews, meta-theoretical analysis, ontology engineering, design science methods, and user-driven innovation in AI projects. He has authored three books and over 100 academic papers, with his work receiving multiple accolades, including the Herbert A. Simon Design Science Award. Beyond academia, Roland has contributed to innovation as a founder and advisory board member of several startups and holds a patent in machine learning.

**Veda C. Storey** [0000-0002-8753-1553] is a Distinguished University Professor and the Tull Professor of Computer Information Systems and professor of computer science at the J. Mack Robinson College of Business, Georgia State University. Her research interests are in data management, conceptual modeling, and design science. She is particularly interested in the assessment of the impact of new technologies on business and society from a data management perspective. Dr. Storey is a member of the AIS College of Senior Scholars and the steering committee of the International Conference of Conceptual Modeling. She is a recipient of the Peter P. Chen Award, an ER Fellow, an AIS Fellow, and an INFORMS Fellow.

**Jeffrey Parsons** [0000-0002-4819-2801] is University Research Professor and Professor of Information Systems in the Faculty of Business Administration at Memorial University of Newfoundland in Canada. His research interests focus on how to better represent human conceptualizations of the world in data. His work on this and related topics has been published in several disciplines. Jeff's research has been recognized in several ways, including *MISQ* paper of the year (2019), AIS Senior Scholars Paper Award (2020), and the INFORMS Design Science Award (2014). He is a Fellow of the Association for Information Systems, Distinguished Research Fellow from TU Dresden, Schoeller Senior Fellow, and an ER Fellow. He has served as a senior editor at the MIS Quarterly and is current a senior editor at Information Systems Research.

**Debra Vander Meer** [0000-0002-5930-6667] is a professor of information systems at Florida International University (FIU). Her current research interests focus on applying concepts from computer science and information systems to real-world problems. Her work appears in premier journals and conferences in these areas. Prior to joining FIU, she spent a decade working in industry, including roles in a venture-funded startup. She holds a bachelor's degree from Georgetown University, a Master of Science in Management Information Systems from the University of Arizona, and a doctoral degree in Computer Science from the Georgia Institute of Technology.

**Dirk S. Hovorka** [0000-0001-7049-5617] is a Professor in the Business Information Systems discipline at the University of Sydney, Australia. He received his PhD in Information Systems from the University of Colorado and holds an MS in Interdisciplinary Telecommunications and an MS in Geology. His current research explores how scientific and societal practices bring forth possible 'worlds' through theory, design, and conceptualizations of futures. His research focuses on *speculative approaches* to 'knowing' regarding technology, society, and biophysical environments in recognition that the future is implacably resistant to empirical knowing, and that unprecedented sociotechnical change renders our knowledge about the past less indicative of future states. Dirk is Senior Editor at JAIS (Research Perspectives), Editorial Board Member of ISR, and Co-chair of the HICSS "Informing Research: Engaging with Futures" mini-track and ECIS "Future as a Site of Inquiry" track. He is an honoree of the 2018 Beta Gamma Sigma Professor of the Year, the Wayne Lonergan Award for Outstanding Teaching (2018), the Dean's Award for Teaching (2022) and has numerous Teaching Citations. Dirk is a co-author of the AIS 2011 Best Paper "Secondary Design: A Case of Behavioral Design Research."

## Appendix A. Overall Usage of Design Science Validity Types

We examined the extent to which the Design Science Validity Framework, translated into the existing 70 validity definitions sorted into our framework in the right column of Table 7, were used in 199 design science papers published in the AIS Senior Scholars' Basket of Eight journals between April 2004 and December 2017. Each paper was examined at the sentence level and compared to our list of validity types, using regular expressions at the word level. The first author read all resulting sentences and excluded search results where their regular expressions yielded excessive false positives. For example, terms such as *accuracy*, *completeness*, *precision*, and *recall* have specific, but polysemous, meanings in data quality and machine learning research. The remaining terms yielded tens of thousands of sentences, so the reported results are conservative estimates.

We identified the number of times a validity corresponding to one in the framework (Table 7), was used at least once within a paper, organized by year and validity type. We concluded that design science papers do not use the same validity terms used in other disciplines, except some used to describe metrics from the confusion matrix. The validity names associated with the characteristic validity types were seldom employed. Even within the highly used category of efficacy validity types, most validity-related terms discussed stemmed from confusion matrix measures in machine learning. Exceptions were in the use of *method characteristic validity types*, the next most used validity types found in the design science literature.

We found a lack of consistency and a lack of actual use of validity terms in design science papers, which implies a significant opportunity for improved communication of evaluation by greater consistency in language around explicit validity claims. Once researchers commit to shared validity norms, communication and reporting consistency should improve.